\documentclass[submission,copyright,creativecommons]{eptcs}
\providecommand{\event}{HCVS 2023} % Name of the event you are submitting to
\usepackage{breakurl}             % Not needed if you use pdflatex only.
\usepackage{multirow}
\usepackage{amssymb}
\usepackage{xspace}
\usepackage[noabbrev,capitalize]{cleveref}
\usepackage{hhline}
\usepackage[htt]{hyphenat}
\usepackage{wrapfig}
\usepackage[table,x11names]{xcolor}
\usepackage{orcidlink}

\title{CHC-COMP 2023: Competition Report}
\author{
Emanuele De Angelis\thanks{
The author is member of the INdAM Research Group GNCS.
}
\institute{IASI-CNR, Italy}
\email{emanuele.deangelis@iasi.cnr.it}
\and
Hari Govind V K
\institute{University of Waterloo, Canada}
\email{hgvk94@gmail.com}
}
\def\titlerunning{CHC-COMP 2023: Competition Report}
\def\authorrunning{Emanuele De Angelis and Hari Govind V K}
\newcommand{\spacer}{Spacer}
\newcommand{\eldarica}{Eldarica}
\newcommand{\golem}{Golem}
\newcommand{\unihorn}{U. Unihorn}
\newcommand{\treeautomizer}{U. TreeAutomizer}
\newcommand{\loat}{LoAT}
\newcommand{\thetafix}{Theta}

\begin{document}
% --------------------------------------------------------
\maketitle
% --------------------------------------------------------
\begin{abstract}
CHC-COMP 2023 is the sixth edition of the Competition of Solvers for Constrained Horn Clauses.
The competition was run in April 2023 and the results were presented at the 10th Workshop on Horn
Clauses for Verification and Synthesis held in Paris, France, on April 23, 2023.
This edition featured seven solvers (six competing and one hors concours) and six tracks, each of which dealing with a class of clauses.
This report describes the organization of CHC-COMP 2023 and presents its results.
\end{abstract}
% --------------------------------------------------------
\section{Introduction}
\textit{Constrained Horn Clauses} (CHCs) are a class of first-order logic formulas where the Horn clause format is extended with \textit{constraints}, that is, formulas of an arbitrary, possibly non-Horn, background theory (such as linear integer arithmetic, arrays, and algebraic data types).

CHCs have gained popularity as a formalism well suited for automatic program verification~\cite{Gr&12,BGMR2015,DFGHPP2021}).
Indeed, the last decade has seen impressive progress in the development of solvers for CHCs (CHC solvers), which can now be effectively used as back-end tools for program verification due to their ability to solve satisfiability problems dealing with a variety of background theories. 
A non-exhaustive list of solvers includes:
ADTInd~\cite{YangFG19},
ADTRem~\cite{De&22a},
Eldarica~\cite{FMCAD2018HojjatRummer},
FreqHorn~\cite{FedyukovichZG18},
Golem~\cite{Blicha_2022},
HSF~\cite{Gr&12},
PCSat~\cite{Unno&21},
RAHFT~\cite{Ka&16},
RInGen~\cite{kostyukov2021finite},
SPACER~\cite{Komuravelli2016},
Ultimate TreeAutomizer~\cite{TreeAutomizer}, and
VeriMAP~\cite{De&14b}.

CHC-COMP is an annual competition that aims to evaluate state-of-the-art CHC solvers on realistic and publicly available benchmarks;
it is open to proposals and contributions from users and developers of CHC solvers, as well as researchers working in the field of CHC solving foundations and its applications.

CHC-COMP 2023\footnote{\url{https://chc-comp.github.io/}} is the 6th edition of the CHC-COMP, affiliated with the 10th Workshop on Horn Clauses for Verification and Synthesis (HCVS 2023\footnote{\url{https://www.sci.unich.it/hcvs23/}}) held in Paris, France, on April 23, 2023.
The deadline for submitting candidate benchmarks was March 24, 2023. 
The deadlines for submitting tools for the test (optional) and the competition runs were 31 March and 7 April 2023, respectively.
The competition was run in the subsequent two weeks, and the results were announced at HCVS 2023.
CHC-COMP 2023 featured 7 solvers (6 competing solvers and one hors concours), and 6 tracks, each of which dealing with a class of clauses
consisting of linear and nonlinear CHCs with constraints over linear integer arithmetic, arrays, non-recursive/recursive algebraic data types, and a few combinations thereof.

This report is structured as follows. 
Section~\ref{sec:D&O} presents the competition tracks, the technical resources used to run the competition, and the evaluation model adopted to rank the solvers.
Section~\ref{sec:benchmarks} presents the inventory of benchmarks and how the candidate benchmarks have been processed and selected for the competition runs.
Sections~\ref{sec:solvers} and~\ref{sec:results} present the tools submitted to CHC-COMP 2023 and the results of the competition, respectively. 
Section~\ref{sec:conclusions} presents some closing remarks from the organizers and participants of CHC-COMP 2023.
Section~\ref{sec:tools} collects the tool descriptions contributed by the participants.
Finally, Appendix~\ref{app:detres} includes the tables with the detailed results about the competition runs.

\subsection*{Acknowledgements} 
We would also like to thank the HCVS 2023 Program Chairs, David Monniaux and Jose F. Morales, for hosting the competition this year as well, and all the HCVS attendees for the fruitful discussion we had after the presentation of the CHC-COMP report.
A special thanks goes to Hossein Hojjat for presenting CHC-COMP 2023 at TOOLympics 2023\footnote{\url{https://tacas.info/toolympics2023.php}}. 

CHC-COMP 2023 heavily built on the infrastructure developed by the organizers of the previous editions, that is, Grigory Fedyukovich, Arie Gurfinkel, and Philipp R\"{u}mmer, which also includes the contributions from Nikolaj Bj{\o}rner, Adrien Champion, and Dejan Jovanovic.

We are also extremely grateful to StarExec\footnote{\url{https://www.starexec.org/}}~\cite{starexec} that continues to support the CHC community by providing the CHC-COMP the computing resources to run the competition. 
In particular, we would like to thank Aaron Stump for helping us in accessing and using the StarExec services.

% ----------------------------------------------------------------
\section{Design and Organization}
\label{sec:D&O}

This section presents 
(i) the competition tracks,
(ii) the technical resources used to run the solvers,
(iii) the characteristics of the test and the competition runs, and
(iv) the evaluation model used to rank the solvers in each track.

\newcommand{\LIAlin}{LIA-lin\xspace}
\newcommand{\LIAnonlin}{LIA-nonlin\xspace}
\newcommand{\LIAarrlin}{LIA-lin-Arrays\xspace}
\newcommand{\LIAarrnonlin}{LIA-nonlin-Arrays\xspace}
\newcommand{\LRATSlin}{LRA-TS\xspace}
\newcommand{\LRATSpar}{LRA-TS-par\xspace}
\newcommand{\ADTarrnonlin}{LIA-nonlin-Arrays-nonrecADT\xspace}
\newcommand{\ADTLIAnonlin}{ADT-LIA-nonlin\xspace}
\newcommand{\ADTnonlin}{ADT-nonlin\xspace}

\subsection{Tracks}
CHC-COMP is organized in tracks, each of which deals with a class of CHCs. 
CHCs are classified according to the following features:
(i) the background theory of the constraints, and 
(ii) the number of \textit{uninterpreted atoms} (that is, atoms whose predicate symbols do not belong to the background theory) occurring in the premises of clauses.
A clause with at most one uninterpreted atom in the premise is said to be \textit{linear}, and \textit{nonlinear} otherwise.

Solvers participating in the CHC-COMP 2023 could enter the competition in six tracks (one track was introduced in this edition, that is, \ADTLIAnonlin, while the remaining tracks were inherited from the previous edition)\footnote{No solver requiring the syntactic restriction on the form of the clauses included in the \LRATSlin track has been submitted in last two editions. Hence, as proposed in~\cite{chccomp21,chccomp22}, the \LRATSlin and \LRATSpar tracks have been discontinued.
Similarly, by considering recent advances in solving techniques for CHCs including algebraic data types, the syntactic restriction on the constraints of the CHCs in the \ADTnonlin track, which requires to have all theory symbols encoded as ADTs (called ``pure ADT'' problems in~\cite{chccomp21}), was no longer needed. Hence, the track has been discontinued and replaced with a more general track combining LIA and ADTs (that is, \ADTLIAnonlin).}:

\begin{enumerate}

\item \textbf{\LIAlin}: Linear Integer Arithmetic - linear clauses

\item \textbf{\LIAnonlin}: Linear Integer Arithmetic - nonlinear clauses

\item \textbf{\LIAarrlin}: Linear Integer Arithmetic \& Arrays - linear clauses 

\item \textbf{\LIAarrnonlin}: Linear Integer Arithmetic \& Arrays - nonlinear clauses 

\item \textbf{\ADTarrnonlin}: Linear Integer Arithmetic \& Arrays \& nonrecursive Algebraic Data Types - nonlinear clauses

\item \textbf{\ADTLIAnonlin}: Algebraic Data Types \& Linear Integer Arithmetic - nonlinear clauses

\end{enumerate}

In addition to the theories occurring in the above list (Linear Integer Arithmetic, Arrays, nonrecursive/recursive Algebraic Data Types, and combinations thereof), benchmarks in all tracks can also make use of the Bool theory.

Finally, in LIA constraints we allow the syntactic appearance of 
the function symbols $*$, \textit{div}, \textit{mod}, and \textit{abs}. 
If these operations do appear, the benchmark is included/excluded from the set of LIA benchmarks according to the following rules:
(i) if the second argument of any \textit{div} and \textit{mod} 
operation is not a constant term, the benchmark is excluded;
(ii) if there is more than one non-constant term in any $*$ operation, the benchmark is excluded;
(iii) otherwise, the operations are considered semantically linear and the benchmark is included.

\subsection{Technical Resources}
CHC-COMP 2023 was run, as well as in the previous editions, on the StarExec platform, but using different technical resources\cite{chccomp22}. 
StarExec made available to the CHC community a queue, called \texttt{chcseq.q}, consisting of 20 brand new nodes equipped with 
Intel(R) Xeon(R) Gold 6334 CPUs.
The detailed specification of the machine is available on the StarExec webpage\footnote{\url{https://www.starexec.org/starexec/public/machine-specs.txt}}. 

%New specs (see Aaron's email):
% \begin{verbatim}
% [root@z001 ~]# cat /etc/centos-release
% CentOS Linux release 7.9.2009 (Core)
% [root@z001 ~]# uname -r
% 3.10.0-1160.62.1.el7.x86_64
% [root@z001 ~]# cat /proc/cpuinfo  | egrep "^model name|^cache size" | head -2
% model name	: Intel(R) Xeon(R) Gold 6334 CPU @ 3.60GHz
% cache size	: 18432 KB
% [root@z001 ~]# cat /proc/meminfo  | grep MemTotal
% MemTotal:       263588804 kB
% [root@z001 ~]# rpm -qa | egrep "^glibc-[0-9]|^gcc-[0-9]"
% glibc-2.17-325.el7_9.i686
% gcc-4.8.5-44.el7.x86_64
% glibc-2.17-325.el7_9.x86_64
% \end{verbatim}

\subsection{Test and Competition Runs}

CHC solvers are evaluated by performing a \textit{test} run
and a \textit{competition} run on the StarExec platform.
A run involves submitting jobs to StarExec, that is, 
collections of 
$\langle$solver-configuration, benchmark$\rangle$
pairs.

The \textit{test} run is used by the participants to 
get acquainted with the StarExec platform and test out their pre-submissions. 
Submitting a solver for \textit{test} runs is optional.  During this test phase, 
the organizers contact the participants 
if they find any issues with their submission so that the participants can
fix it before their final submission. 
The participants are given a week in between the 
\textit{test} and \textit{competition} runs. In the \textit{test} runs, 
a small set of randomly selected benchmarks
is used, and each job is limited to 600s CPU time, 600s 
wall-clock time, and 64GB memory.
% In test runs, the (optional) pre-submissions 
% of the solvers are evaluated to check their configurations 
% and identify possible inconsistencies.
% In these runs a small set of randomly selected benchmarks
% is used, and each job is limited to 600s CPU time, 600s 
% wall-clock time, and 64GB memory.

In \textit{competition} runs, the final submissions
of the solvers are evaluated to determine the outcome 
of the competition, that is, to rank the solvers that
entered the competition. 
In these runs each job is limited to 1800s CPU time, 
1800s wall-clock time, and 64GB memory.

Sometimes, the competition benchmarks expose soundness bugs in solvers. We catch these bugs if two solvers disagree on the satisfiability of a benchmark. At CHC-COMP, we keep things friendly by informing the participants about the inconsistency and giving them the benchmark to reproduce the issue. If we have time, we even give them a chance to fix the issue and resubmit their tool. If not, we disqualify the tool from the track.

The data gathered from the `job information' CSV files produced by StarExec in the competition runs are used to rank the solvers.
All `job information' CSV files of the CHC-COMP 2023 runs are available on the StarExec space \texttt{CHC/CHC-COMP/CHC-COMP-23}\footnote{\url{https://www.starexec.org/starexec/secure/explore/spaces.jsp?id=538944}}.

\subsection{Evaluation of the Competition Runs}
\label{subsub:eval}

The competing solvers were evaluated using the same approach as the 2022 edition~\cite{chccomp22}.

The evaluation of the competition runs were done using the \texttt{summarize.py} script available at \url{https://github.com/chc-comp/scripts}; the script takes as input the `job information' CSV file produced by StarExec at job completion, and produces a ranking of the solvers.

The ranking of solvers in each track is based on the score obtained by the solvers in the competition run for a track.
The score is computed on the basis of the results provided by the solver on the benchmarks for that track.
The result can be \textit{sat}, \textit{unsat}, or \textit{unknown} (which includes solvers giving up, running out of resources, or crashing), and the score given by the number of \textit{sat} or \textit{unsat} results.
In the case of ex-aequo, the ranking is determined by using the CPU time, which is the total CPU time needed by a solver to produce the results.

The tables in Appendix~\ref{app:detres} also report in column `\#unique' the number of \textit{sat} or \textit{unsat} results produced by a solver for benchmarks for which all other solvers returned \textit{unknown}.
The `job information' files also include data about the space and memory consumption, which we consider less relevant and therefore are not reported in the tables (see also the CHC-COMP 2021 and CHC-COMP 2022 reports~\cite{chccomp21,chccomp22}).

% ----------------------------------------------------------------
\section{Benchmarks}
\label{sec:benchmarks}

\subsection{Format}
CHC-COMP accepts benchmarks in the SMT-LIB 2.6 
format~\cite{BarFT-SMTLIB}.
All benchmarks have to conform to the format described
at \url{https://chc-comp.github.io/format.html}. This year, 
we updated the format to allow the declaration of ADTs using the \texttt{declare-datatypes} command. We support ADTs with any number of constructors and selectors as long as they are not parametric.
Conformance is checked using the \texttt{format.py} script
available at \url{https://github.com/chc-comp/scripts}.

\subsection{Inventory}
All benchmarks used for the competition are selected 
from repositories under \url{https://github.com/chc-comp}.
Anyone can contribute benchmarks to this repository.
This year, we got several new benchmarks for the track \textbf{\ADTLIAnonlin}. Table~\ref{tab:bench_summary} summarizes 
the number of benchmarks and unique benchmarks available
in each repository. The organizers pick a subset of all
available benchmarks for each year's competition.
In the rest of this section, we explain the steps in
this selection.

\subsection{Processing Benchmarks}
All benchmarks are processed using the \texttt{format.py}
script, which is available at \url{https://github.com/chc-comp/scripts}. The command line for invoking the script is

\noindent
\begin{verbatim}
   > python3.9 format.py --out-dir <out-dir> --merge_queries True <smt-file> 
\end{verbatim}

\noindent
The script attempts to put benchmark \texttt{<smt-file>} into CHC-COMP format. The \texttt{merge\_queries} option merges multiple queries into a single query as discussed in previous editions of CHC-COMP~\cite{chccomp21}. In previous competitions, this script was not used in tracks containing ADTs because it did not print ADTs. This year, we updated the script to support printing ADTs in the SMT-LIB format using the \texttt{declare-datatypes} command. When printing ADTs are grouped as follows: if a constructor of ADT type $a$ takes an argument of type ADT $b$, both $a$ and $b$ are grouped together. All ADTs in a group are declared together inside the same \texttt{declare-dataypes} command.

After formatting, benchmarks are categorized into one of  the 6 competition tracks: LIA-lin, LIA-nonlin, LIA-lin-Arrays, LIA-nonlin-Arrays, ADT-LIA-nonlin, and LIA-nonlin-Arrays-nonrecADT. The scripts for categorizing benchmarks are available at~\url{https://github.com/chc-comp/chc-tools/tree/master/format-checker}. This year, we added support for ADT tracks in the categorizing script. The script now checks for proper declaration of ADTs and proper usage of constructors, selectors, and recognizers. However, it does not check if a given ADT is recursive or not. Therefore, for the LIA-nonlin-Arrays-nonrecADT track, we manually verified that all ADTs are non-recursive. Benchmarks that could not be put in CHC-COMP compliant format and benchmarks that could not be categorized into any tracks are not used for the competition.

\bigskip
\begin{table}[!ht]
\centering
\begin{tabular}{
@{\hspace{2pt}}l@{\hspace{6pt}}
@{\hspace{2pt}}c@{\hspace{2pt}}
@{\hspace{2pt}}c@{\hspace{2pt}}
@{\hspace{2pt}}c@{\hspace{2pt}}
@{\hspace{2pt}}c@{\hspace{2pt}}
@{\hspace{2pt}}c@{\hspace{2pt}}
@{\hspace{2pt}}c@{\hspace{2pt}}}
\hline
Repository 
&  \multirow{1}{1.8cm}{LIA-\\lin}     
&  \multirow{2}{1.8cm}{LIA-\\nonlin}   
&  \multirow{3}{1.8cm}{LIA-\\lin-\\Arrays} 
&  \multirow{3}{1.9cm}{LIA-\\nonlin-\\Arrays} 
&  \multirow{4}{2.1cm}{LIA-\\nonlin-\\Arrays-\\nonrecADT}
&  \multirow{3}{1.9cm}{ADT-\\LIA-\\nonlin}\\
                     &               &            &            &         & &\\
                     &               &            &            &         & &\\
                     &               &            &            &         & &\\\hline\hline
adtrem~(\textit{new})&               &            &            &         & & 251/247\\\hline
aeval                &  54/54        &            &            &         & &\\\hline
aeval-unsafe         &  54/54        &            &            &         & &\\\hline
chc-comp19           &               &            & 290/290    &         & &\\\hline
eldarica-misc        &  149/136      &  69/66     &            &         & &\\\hline
extra-small-lia      &  55/55        &            &            &         & &\\\hline
hcai                 &  101/87       &  133/131   & 39/39      &   25/25 & &\\\hline
hopv                 &  49/48        &   68/67    &            &         & &\\\hline
jayhorn              &  75/73        & 7325/7224  &            &         & &\\\hline
kind2                &               &  851/736   &            &         & &\\\hline
ldv-ant-med          &               &            & 10/10      & 342/342 & &\\\hline
ldv-arrays           &               &            & 3/2        & 821/546 & &\\\hline
llreve               &  66/66        &      59/57 & 31/31      &         & &\\\hline
quic3                &               &            & 43/43      &         & &\\\hline
rust-horn~(\textit{new}) & 11/11     &      6/6   &            &         & & 56/56\\\hline            
seahorn              &  3379/2812    &     68/66  &            &         & &\\\hline
solidity             &               &            &            &         & 2200/2174 & \\\hline
sv-comp              &  3150/2930    &  1643/1169 & 79/73      & 856/780 & &\\\hline
synth/nay-horn       &               &    119/114 &            &         & &\\\hline
synth/semgus         &               &            &            & 5371/4839 & &\\\hline
tip-adt-lia~(\textit{new}) &         &            &            &         & & 320/320\\\hline
tricera              &  405/405      &  4/4       &            &         & &\\\hline
tricera/adt-arrays   &               &            &            &         & 156/156 &\\\hline
ultimate             &               &  8/8       &            &  23/23  & &\\\hline
vmt                  &  906/803      &            &            &         & &\\\hline\hline
total/{\bf unique}    & 8454/{\bf 7534} & 10353/{\bf 9648} 
                   &   495/{\bf 488} & 7438/{\bf 6555} 
                   &   2356/{\bf 2330 } & 627/{\bf 623} \\\hline
\end{tabular}
\caption{Summary of benchmarks (total/unique).}
\label{tab:bench_summary}
\end{table}

\clearpage
% \subsection{Processing benchmarks with Algebraic Data Types and Reals}
% For benchmarks containing either ADTs or Reals, no processing is done. All benchmarks submitted to the ADT-nonlin track were already processed using the \textsc{RInGen} tool~\cite{kostyukov2021finite} to encode all theory symbols using ADTs. The benchmarks submitted to the LRA-TS and LIA-nonlin-Arrays-nonrecADT track were already in compliance with the CHC-COMP format.

\subsection{Rating and Selection}
\label{subsub:selBench}
This section describes the procedure used to select benchmarks for the
competition.

We picked all unique benchmarks in the \LIAarrlin track because of the scarcity 
of available benchmarks. In all other tracks, we followed a procedure similar to the
past editions of the competition aiming at selecting a representative subset
of the available benchmarks.
In particular, we estimated how ``easy" the benchmarks were and picked
a mix of ``easy" and ``hard" instances.
We say that a benchmark in a track is ``easy" if it is solved by both
the winner and the runner-up solvers in the corresponding track in 
CHC-COMP 2022, within a small time interval~(30s).

Each benchmark was rated A/B/C based on how difficult the winner and the runner-up solvers found them. A rating of ``A" is given if both solvers solved the benchmark, ``B" if only one solver solved it, ``C" if neither solved it, within the set timeout~(30s). We ran all solvers using the same binaries and configurations submitted for CHC-COMP 2022.

Once we labeled each benchmark from a repository $r$, we decided the maximum number of instances, $N_r$, to take from the repository. $N_r$ number was decided based on the total number of unique benchmarks and our knowledge about the benchmarks in repository $r$. 

We picked at most $0.2\cdot N_r$ benchmarks with rating A. Then, we picked at most $0.4\cdot N_r$ benchmarks with rating B; namely, $0.2\cdot N_r$ from those solved only by the winner solver and $0.2\cdot N_r$ from those solved only by the runner-up solver. Finally, we picked at most $0.4\cdot N_r$ benchmarks with rating C. If we did not find enough benchmarks with rating A, we picked the rest of the benchmarks with rating B (equally from those solved only by the winner and the runner-up). If we did not find enough benchmarks with rating B, we pick the remaining benchmarks from rating C.

This way, we obtained a mix of ``easy" and ``hard" benchmarks with a bias towards benchmarks that were not easily solved by either of the best solvers from the previous year's competition. The number of instances with each rating is given in~\cref{tab:bench_rating_1,tab:bench_rating_2}. The number of instances picked from each repository is given in Table~\ref{tab:inst_picked}. To pick \texttt{<num>} benchmarks of rating \texttt{<Y>}, we used the command

\begin{verbatim}
    > cat <rating-Y-benchmark-list> | sort -R | head -n <num>
\end{verbatim}

We were unable to run more than one solver for tracks containing ADTs~(\ADTLIAnonlin, \ADTarrnonlin). Only 3 solvers participated in tracks containing ADTs in CHC-COMP 2022: Spacer, Eldarica, and RInGen. RInGen does not support theories other than ADTs. The version of Eldarica submitted to CHC-COMP 2022 does not support the updated format of CHC-COMP 2023. Specifically, this version of Eldarica does not support the SMT-LIB syntax for recognizers\footnote{since then, Eldarica has been updated to support recognizers. E.g. Eldarica v2.0.9 that participated in CHC-COMP 2023.}. Therefore, we were limited to using just one solver, Spacer, to select benchmarks for tracks containing ADTs. For each repository $r$, we decided a maximum number of instances $N_r$, ran Spacer on all benchmarks with the same timeout~(30s), and picked $0.4\cdot N_r$ benchmarks that Spacer solved (column $B$ in Table~\ref{tab:bench_rating_2}) and $0.6\cdot N_r$ benchmarks that Spacer did not solve (column $C$ in Table~\ref{tab:bench_rating_2}).

The final set of benchmarks selected for CHC-COMP 2023 can be found in the github repository
\url{https://github.com/chc-comp/chc-comp23-benchmarks}, and on StarExec in the public space \texttt{CHC/CHC-COMP-23/CHC-COMP-23-competition-runs}\footnote{\url{https://www.starexec.org/starexec/secure/explore/spaces.jsp?id=538230}}.

% ------------------------------------------------------------------
\begin{table}[!ht]
\centering
\begin{tabular}{@{\hspace{2pt}}l                     @{\hspace{8pt}}
  r@{\hspace{6pt}}r@{\hspace{0pt}}r@{\hspace{6pt}}r@{\hspace{18pt}}
  r@{\hspace{6pt}}r@{\hspace{0pt}}r@{\hspace{6pt}}r@{\hspace{18pt}}
  r@{\hspace{6pt}}r@{\hspace{0pt}}r@{\hspace{6pt}}r}
\hline
&  \multicolumn{4}{c}{\LIAlin~~~~~}      
&  \multicolumn{4}{c}{\LIAnonlin~~~~~}  
&  \multicolumn{4}{c}{\LIAarrnonlin}\\
%Repository           &  \#A & \#${\rm B_1}$&\#${\rm B_2}$ & \#C&   \#A&  \#${\rm B_1}$&\#${\rm B_2}$& \#C&   \#A&\#${\rm B_1}$&\#${\rm B_2}$& \#C \\\hline\hline
Repository           &  \#A & \multicolumn{2}{c}{~\#B} & \#C 
                     &  \#A & \multicolumn{2}{c}{~~\#B} & \#C
                     &  \#A & \multicolumn{2}{c}{~~\#B} & \#C \\[-2pt]
                     &   & (\textit{w}) & (\textit{r}) & 
                     &   & (\textit{w}) & (\textit{r}) & 
                     &   & (\textit{w}) & (\textit{r}) &  \\[1pt]\hline
aeval                &   12&  9&  4&  29&      &     &  &     &     &     &  &     \\\hline
aeval-unsafe         &   17&  0& 12&  25&      &     &  &     &     &     &  &     \\\hline
eldarica-misc        &  120&  5&  9&   2&    39&   13& 0&   14&     &     &  &     \\\hline
extra-small-lia      &   30& 13&  8&   4&      &     &  &     &     &     &  &     \\\hline
hcai                 &   82&  1&  3&   1&   123&    0& 5&    3&   17&    3& 0&    5\\\hline
hopv                 &   48&  0&  0&   0&    57&    3& 5&    2&     &     &  &     \\\hline
jayhorn              &   73&  0&  0&   0&  3712& 2275& 1& 1236&     &     &  &     \\\hline
kind2                &     &   &   &    &   650&   70& 0&   16&     &     &  &     \\\hline
ldv-ant-med          &     &   &   &    &      &     &  &     &    0&  128& 0&  214\\\hline
ldv-arrays           &     &   &   &    &      &     &  &     &    7&  195& 0&  344\\\hline
llreve               &   61&  0&  5&   0&    48&    4& 2&    3&     &     &  &     \\\hline
rust-horn            &   10&  1&  0&   0&     5&    0& 0&    1&     &     &  &     \\\hline
seahorn              & 2089& 65& 69& 589&    60&    1& 2&    3&     &     &  &     \\\hline
sv-comp              & 2854&  1& 74&   1&  1117&   40& 4&    8&  310&  330& 7&  133\\\hline
synth/nay-horn       &     &   &   &    &    70&   20& 4&   20&     &     &  &     \\\hline
synth/semgus         &     &   &   &    &      &     &  &     &  737& 2254& 4& 1844\\\hline
tricera/svcomp20     &   43&  7&  4& 351&     4&    0& 0&    0&     &     &  &     \\\hline
ultimate             &     &   &   &    &     0&    1& 0&    7&    0&    0& 0&   23\\\hline
vmt                  &  711& 31&  7&  54&      &     &  &     &     &     &  &     \\\hline\hline
{\bf total }         & ~6150& ~~133& ~~195& ~1056&  ~5885& ~2427& ~~~~23& ~1313& ~1071& ~2910& ~~~~11& 2563\\\hline
\end{tabular}
\caption{The number of unique benchmarks with ratings A/B/C - Tracks: \LIAlin, \LIAnonlin, and \LIAarrnonlin.
B-rated benchmarks are reported in two sub-columns: 
(\textit{w}) benchmarks solved only by the CHC-COMP 2022 winner, and (\textit{r}) 
benchmarks solved only by the CHC-COMP 2022 runner-up solver.
}
\label{tab:bench_rating_1}
\end{table}
% ----------------------------------------------------------------------
\begin{table}[!ht]
\centering
\begin{tabular}{l  @{\hspace{15pt}}
  r@{\hspace{5pt}}r@{\hspace{10pt}}
  r@{\hspace{5pt}}r@{\hspace{5pt}}} \hline
           &\multicolumn{2}{l}{LIA-nonlin-}      & \multicolumn{2}{l}{ADT-} \\
           &\multicolumn{2}{l}{Arrays-}          & \multicolumn{2}{l}{LIA-nonlin} \\
           &\multicolumn{2}{l}{nonrecADT} & \\
Repository         & \#B &  \#C &  \#B &  \#C\\\hline\hline
adtrem             &     &      &    86&  161\\\hline
rust-horn          &     &      &    43&   13\\\hline
solidity           & 2109&    65&      &     \\\hline
tip-adt-lia        &     &      &    39&  281\\\hline
tricera/adt-arrays &   65&    91&      &     \\\hline\hline
{\bf total }       & ~2174&~~156& ~~168&~~455\\\hline
\end{tabular}
\caption{The number of unique benchmarks with ratings B/C -- Tracks: \ADTnonlin, and \ADTarrnonlin.}
\label{tab:bench_rating_2}
\end{table}

\begin{table}[!htbp]
\centering
\begin{tabular}{
@{\hspace{2pt}}l@{\hspace{0pt}}
@{\hspace{3pt}}r@{\hspace{3pt}}
@{\hspace{3pt}}r@{\hspace{3pt}}
@{\hspace{3pt}}r@{\hspace{3pt}}
@{\hspace{3pt}}r@{\hspace{3pt}}
@{\hspace{3pt}}r@{\hspace{3pt}}}
\hline
Repository 
&  \multirow{1}{1.4cm}{\LIAlin}     
&  \multirow{2}{1.4cm}{\LIAnonlin}   
&  \multirow{3}{1.4cm}{\LIAarrnonlin} 
&  \multirow{4}{1.7cm}{\ADTarrnonlin}
&  \multirow{2}{1.4cm}{\ADTLIAnonlin}\\
                     &        &         &         &         & \\
                     &       &          &         &         & \\
                     &       &          &         &         & \\\hline\hline
adtrem               &       &          &         &         & 125/125\\\hline
aeval                & 30/30 &          &         &         & \\\hline
aeval-unsafe         & 30/30 &          &         &         & \\\hline
eldarica-misc        & 45/25 &   30/26  &         &         & \\\hline
extra-small-lia      & 30/22 &          &         &         & \\\hline
hcai                 & 45/14 &   60/20  &  15/11  &         & \\\hline
hopv                 &  30/6 &   30/16  &         &         & \\\hline
jayhorn              &  30/6 &  180/180 &         &         & \\\hline
kind2                &       &   90/52  &         &         & \\\hline
ldv-ant-med          &       &          &  60/60  &         & \\\hline
ldv-arrays           &       &          &  90/90  &         & \\\hline
llreve               & 30/11 &   45/18  &         &         & \\\hline
rust-horn            &       &          &         &         & 28/18\\\hline
seahorn              & 90/90 &   45/15  &         &         & \\\hline
solidity             &       &          &         & 312/127 & \\\hline
sv-comp              & 90/38 &   90/48  & 135/135 &         &\\\hline
synth/nay-horn       &       &   60/48  &         &         &\\\hline
synth/semgus         &       &          & 135/135 &         &\\\hline
tip-adt-lia          &       &          &         &         & 160/160\\\hline
tricera/svcomp20     & 60/60 &     3/0  &         &         & \\\hline
tricera/adt-arrays   &       &          &         & 156/122 &  \\\hline
ultimate             &       &     6/5  &  15/15  &         &\\\hline
vmt                  & 90/90 &          &         &         &\\\hline\hline
{\bf total }& 600/{\bf 422 } &  639/{\bf 428 } & 450/{\bf 446 } & 468/{\bf 249 } & 313/{\bf 303 }\\\hline

\end{tabular}
\caption{The number of benchmarks to select and the number of selected benchmarks from each repository.}
\label{tab:inst_picked}
\end{table}

% ----------------------------------------------------------------
\clearpage
\section{Solvers}
\label{sec:solvers}

Seven solvers were submitted to CHC-COMP 2023: six competing solvers, and one solver \textit{hors concours} (Spacer is co-developed by Hari Govind V~K who is co-organizing the CHC-COMP 2023.).

Table~\ref{tab:solvers} lists the submitted solvers together with the configurations used to run them on the competition tracks.
Detailed descriptions of the solvers are provided in Section~\ref{sec:tools}. 
The binaries of the solvers are available on the StarExec space
\texttt{CHC/CHC-COMP/CHC-COMP-23-competitions-runs}. 

\begin{table}[h]
    \centering
    \begin{tabular}{
@{\hspace{5pt}}p{2.5cm}|
@{\hspace{5pt}}p{2.1cm}
@{\hspace{5pt}}p{2.2cm}
@{\hspace{5pt}}p{2.1cm}
@{\hspace{5pt}}p{2.1cm}
@{\hspace{5pt}}p{2.1cm}
@{\hspace{5pt}}p{1cm}
}
\hline
                     &            &            &            &         & &\\[-11pt]
\multirow{4}{*}{\textbf{Solver}}
&  \multirow{2}{2cm}{LIA-\\lin}     
&  \multirow{2}{2cm}{LIA-\\nonlin}   
&  \multirow{3}{2cm}{LIA-\\lin-\\Arrays} 
&  \multirow{3}{2cm}{LIA-\\nonlin-\\Arrays} 
&  \multirow{4}{2cm}{LIA-\\nonlin-\\Arrays-\\nonrecADT}
&  \multirow{3}{2cm}{ADT-\\LIA-\\nonlin}\\
                     &       &          &  &         &         & \\
                     &       &          &  &         &         & \\                     
                     &       &          &  &         &         & \\[1pt]\hline\hline
    \textbf{Eldarica}      &
                  \texttt{def} &
                  \texttt{def} &
                  \texttt{def} &
                  \texttt{def} & 
                  \texttt{def} &
                  \texttt{def}\\\hline
    \textbf{Golem}         &
                  \texttt{lia-lin}    &
                  \texttt{lia-nonlin} & 
                  &
                  &
                  &
                  \\\hline
    \textbf{LoAT}          & 
                  \texttt{loat\_horn} &
                  &
                  &
                  &
                  & 
                  \\\hline
    \textbf{Theta}         & 
                  \texttt{fix} &
                  \texttt{fix} &
                  \texttt{fix} &
                  \texttt{fix} &
                               &
                  \\\hline
    \textbf{Ultimate      
    TreeAutomizer} &
                  \texttt{default} &
                  \texttt{default} &
                  \texttt{default} &
                  \texttt{default} &
                  &
                  \\\hline
    \textbf{Ultimate ~~~      
    Unihorn}       &
                  \texttt{default} &
                  \texttt{default} &
                  \texttt{default} &
                  \texttt{default} &
                  &
                  \\\hline
    \textbf{Spacer}        &
                  \texttt{def}            &
                  \texttt{def}         & 
                  \texttt{ARRAYS}    &
                  \texttt{ARRAYS} &
                  \texttt{def}            &
                  \texttt{def}\\\hline
    \end{tabular}
    \caption{Solvers and configurations 
    used in the tracks; an empty entry denotes
    that the solver did not enter the competition 
    in that track.
    The configuration names have been taken as is 
    from solver submissions.}
    \label{tab:solvers}
    
\end{table}

% ----------------------------------------------------------------
\section{Results}
\label{sec:results}

The results of the CHC-COMP 2023 are reported in Table~\ref{tab:results}. 
Detailed results are provided in Appendix~\ref{app:detres}. 
All the data gathered from the execution of the StarExec jobs created for 
the competition run are available on the StarExec space 
\texttt{CHC/CHC-COMP/CHC-COMP-23-competitions-runs}. 

\begin{table}[h]
    \centering
    \begin{tabular}{
@{\hspace{1pt}}p{1.5cm}
@{\hspace{3pt}}p{1.6cm}
@{\hspace{1pt}}p{1.7cm}
@{\hspace{1pt}}p{1.7cm}
@{\hspace{1pt}}p{1.7cm}
@{\hspace{1pt}}p{1.9cm}
@{\hspace{3pt}}p{1.7cm}
@{\hspace{0pt}}
}
\cline{2-7}
& & & & & & \\[-11pt]
&  \multirow{2}{1.5cm}{LIA-\\lin}     
&  \multirow{2}{1.7cm}{LIA-\\nonlin}   
&  \multirow{3}{1.7cm}{LIA-\\lin-\\Arrays} 
&  \multirow{3}{1.7cm}{LIA-\\nonlin-\\Arrays} 
&  \multirow{4}{1.9cm}{LIA-\\nonlin-\\Arrays-\\nonrecADT}
&  \multirow{3}{1.1cm}{ADT-\\LIA-\\nonlin}\\
                     &       &          &  &         &         & \\
                     &       &          &  &         &         & \\                     
                     &       &          &  &         &         & \\[1pt]\hline\hline
    \textbf{Winner} &
        \textbf{Golem}    &
        \textbf{Eldarica} &
        \textbf{Eldarica} & 
        \textbf{Eldarica} &
        \textbf{Eldarica} &
        \textbf{Eldarica} \\\hline
    2nd place &
        Eldarica         &
        Golem            & 
        Theta            &
        Ultimate\newline Unihorn &
        &
        \\\hline
    3rd place & 
        Theta            &
        Ultimate\newline Unihorn &
        Ultimate\newline Unihorn &
        Theta            &
        &
        \\\hline
    \end{tabular}
    \caption{Results of CHC-COMP 2023. Spacer, which entered the competition as hors concours solver, 
    placed in the first position of the \LIAlin, \LIAnonlin, \LIAarrlin, and \LIAarrnonlin tracks.}
    \label{tab:results}
\end{table}

% ----------------------------------------------------------------
\subsection{Observed Issues and Fixes during the Competition runs}

This section describes the issues we have run across when using the tools entered in the competition and how we worked with the teams to overcome them.

\bigskip
\paragraph{Ultimate TreeAutomizer and Ultimate Unihorn}
Due to issues in building a version of Z3 that is able to run on StarExec, the final submission for the competition run of the solvers Ultimate TreeAutomizer and Ultimate Unihorn were completed on 14 April, 2023. 

\bigskip
\paragraph{Theta}
In the competition runs of the \LIAarrnonlin track we detected one inconsistent result:
Theta (Theta-default in Table~\ref{tab:theta-inconsistencies}) reported \textit{unsat} on one benchmark, while other solvers reported \textit{sat}.
The inconsistency was detected on April 14, and we informed the team on the same day by sending them the benchmark on which the issue was detected.
The team submitted an updated version of Theta on April 15.
Due to a configuration problem, the updated version of Theta reported \textit{unknown} on all benchmarks.
We informed the team on April 16, who provided an updated version of the solver (Theta-fix in Table~\ref{tab:theta-inconsistencies}) on the same day.

In the competition runs of the \LIAnonlin track we detected one inconsistent result:
Theta-fix reported \textit{sat}, while other solvers reported \textit{unsat}.
The Theta team was informed on April 19 by sending them the benchmark on which the issue was detected.
The team submitted a fixed version (Thetafix-fix in Table~\ref{tab:theta-inconsistencies}) on April 19 that produced no inconsistent results.

The results presented in this report were produces using the fixed version. 
In Table~\ref{tab:theta-inconsistencies} we report the results before and after the fixes. 

\smallskip
\begin{table}[!h]
    \centering
    \begin{tabular}{p{2.3cm} p{1cm}p{1.4cm} p{1cm}p{1.4cm} p{1cm}p{1.4cm} p{1cm}p{1.4cm} }
    \hline
    \multirow{2}{2.2cm}{Theta version}  
              & \multicolumn{2}{l}{\LIAlin}       &
                \multicolumn{2}{l}{\LIAnonlin}    & 
                \multicolumn{2}{l}{\LIAarrlin}    &
                \multicolumn{2}{l}{\LIAarrnonlin} \\
              & \#\textit{sat} & \#\textit{unsat} &
                \#\textit{sat} & \#\textit{unsat} &
                \#\textit{sat} & \#\textit{unsat} &
                \#\textit{sat} & \#\textit{unsat} \\\hline\hline
    Theta-default  & 129 & 53 & 12 & 21 & 148 & 50 & 52 & 40\\\hline
    Theta-fix      & 121 & 49 &  9 & 20 & 135 & 50 & 45 & 39\\\hline
    Thetafix-fix   & 122 & 48 &  8 & 30 & 134 & 50 & 45 & 40\\\hline
    \end{tabular}
    \caption{Results produced by Theta 
    before and after the fixes.}
    \label{tab:theta-inconsistencies}
\end{table}

% ----------------------------------------------------------------
\clearpage
\section{Conclusions and Final Remarks}
\label{sec:conclusions}

We would like to congratulate the winners of the CHC-COMP 2023 (in alphabetical order): 
\textbf{Eldarica} (winner of the following tracks:
\LIAnonlin,
\LIAarrlin,
\LIAarrnonlin,
\ADTarrnonlin, and \LIAarrnonlin),
and \textbf{Golem} (winner of the \LIAlin track).

\medskip
In organizing this edition of the competition we did our best to address some open issues discussed in the report of the CHC-COMP 2022~\cite{chccomp22}.
In particular, we have replaced the \ADTnonlin track with a more general track dealing with the combined theory of LIA and ADTs (\ADTLIAnonlin), and we have extended the CHC format and the tools for processing and selecting the benchmarks to deal with ADTs.
Moreover, as mentioned in the previous reports~\cite{chccomp21,chccomp22}, we have discontinued the obsolete tracks \LRATSlin and \LRATSpar.
Finally, we have made a small change to the candidate benchmarks rating process by increasing the timeout used to evaluate their ``hardness'' (see Section~\ref{subsub:selBench}). Ideally, we would have run the solvers with the same timeout as used in the competition~(20 minutes). However, there are over 7500 benchmarks to pick from and we expect several timeouts irrespective of the time limit. Hence, for practical reasons, we set the timeout to 30 seconds for all solvers (previous editions had lower values that were dependent on the solver used to rate the benchmarks).

\medskip
Below, we report the still open issues that should be further discussed for future editions, and the proposal for new tracks that emerged from the follow-up discussion we had after the presentation of the competition report at HCVS.

\begin{itemize}
\item \textbf{Validation of results} (also discussed in the previous editions~\cite{chccomp21,chccomp22}).
The ability of solvers to generate a witnesses (models or counter-examples) to support their results is a recurrent request by our community members.
Several solvers have support for generating a witness. However, the witness is used mainly for debugging by the developers and having a common format for them is still a work in progress.
As an additional issue, it is often the case that these witnesses are not for the original CHCs but for those obtained after many layers of pre-processing. Transforming these ``internal" witnesses into a witness for the original problem is also a work in progress.
While reaching a consensus on a common format for their encoding would require a thoughtful discussion involving all members of the CHC community, we could begin, as already proposed in the previous reports, by introducing in the CHC-COMP new tracks where the ability of producing a witness is taken into consideration in the computation of the score. 

\item \textbf{Status of benchmarks} (from~\cite{chccomp22}).
In order to assess the correctness of the result provided by the
solvers, each submitted benchmark should explicitly declare the
expected result of the satisfiability problem.
We propose to use the \texttt{( set-info}
$\langle keyword\rangle$ $\langle \textit{attr-value} \rangle$ 
\texttt{)} command with the \texttt{:status} as \textit{keyword},
and either \texttt{sat} or \texttt{unsat} as \textit{attr-value}.

\item \textbf{Parallel tracks}. (Thanks to \textit{Martin Blicha} for having sent us this note.)
We propose a parallel version for each (or some) of the existing tracks.
Instead of putting a limit on the CPU time, only a limit on the wall-clock time would be imposed in the parallel version.
Parallel tracks can be implemented in two ways: either use the solvers' configuration submitted for the classical tracks, or allow a separate submission for the parallel tracks.

\end{itemize}

Finally, we would to stress once again that \textbf{a bigger set
of benchmarks are needed}. Besides submitting their tools, all
participants are invited to contribute with new benchmarks.

% ----------------------------------------------------------------
\section{Solver Descriptions}
\label{sec:tools}

The tool descriptions in this section were contributed by the
participants, and the copyright on the texts remains with the 
individual authors.

\newcommand{\toolname}[1]{\subsection{#1}}
\newcommand{\toolsubmitter}[2]{\noindent #1\\#2\par\smallskip}
\newcommand{\toolalgorithm}{\paragraph{Algorithm.}}
\newcommand{\toolarchitecture}{\paragraph{Architecture and Implementation.}}
\newcommand{\toolconfiguration}{\paragraph{Configuration in CHC-COMP 2023.}}
\newcommand{\toolnew}{\paragraph{New Features in CHC-COMP 2023.}}
\newcommand{\toollink}[2]{\par\bigskip\noindent\url{#1}\\#2}

\bigskip
%%%%% Edit below

\toolname{Eldarica v2.0.9}
\label{eldchccomp23}

% People submitting the solver, and their affiliation.
% This does not necessarily include all tool authors.

\toolsubmitter{Hossein Hojjat}{University of Tehran, Iran}
\toolsubmitter{Philipp R\"ummer}{University of Regensburg, Germany and
  Uppsala University, Sweden}

\toolalgorithm

Eldarica~\cite{FMCAD2018HojjatRummer} is a Horn solver applying
classical algorithms from model checking: predicate abstraction and
counterexample-guided abstraction refinement (CEGAR).  Eldarica can
solve Horn clauses over linear integer arithmetic, arrays, algebraic
data-types, bit-vectors, and the theory of heaps.  It can process Horn
clauses and programs in a variety of formats, implements sophisticated
algorithms to solve tricky systems of clauses without diverging, and
offers an elegant API for programmatic use.

\toolarchitecture

Eldarica is entirely implemented in Scala, and only depends on Java or
Scala libraries, which implies that Eldarica can be used on any
platform with a JVM. For computing abstractions of systems of Horn
clauses and inferring new predicates, Eldarica invokes the SMT solver
Princess~\cite{princess08} as a library.

\toolconfiguration

Eldarica is in the competition run with the option \verb!-portfolio!,
which enables a simple portfolio mode. Four instances of the solver
are run in parallel, with the following options:
\begin{enumerate}
\item \verb!-splitClauses:0 -abstract:off!,
\item \verb!-splitClauses:1 -abstract:off -stac!,
\item \verb!-splitClauses:1 -abstract:off!,
\item \verb!-splitClauses:1 -abstract:relEqs! (the default options).
\end{enumerate}

\toollink{https://github.com/uuverifiers/eldarica}{BSD licence}

\clearpage
%\newcommand{\toolname}[1]{\subsection{#1}}
%\newcommand{\toolsubmitter}[2]{\noindent #1\\#2\par\smallskip}
%\newcommand{\toolalgorithm}{\paragraph{Algorithm.}}
%\newcommand{\toolarchitecture}{\paragraph{Architecture and Implementation.}}
%\newcommand{\toolconfiguration}{\paragraph{Configuration in CHC-COMP-23.}}
%\newcommand{\toolnew}{\paragraph{New Features in CHC-COMP-23.}}
%\newcommand{\toollink}[2]{\par\bigskip\noindent\url{#1}\\#2}

%%%%% Edit below
%\newcommand{\golem}{\textsc{Golem}}
\toolname{\golem}

% People submitting the solver, and their affiliation.
% This does not necessarily include all tool authors.

\toolsubmitter{Martin Blicha}{Universit\`{a} della Svizzera italiana, Switzerland}
\toolsubmitter{Konstantin Britikov}{Universit\`{a} della Svizzera italiana, Switzerland}

\toolalgorithm
\golem{} is a CHC solver under active development that provides several backend engines implementing various SMT- and interpolation-based model-checking algorithms.
It supports the theory of Linear Real or Integer Arithmetic and it is able to provide witnesses for both satisfiable and unsatisfiable CHC systems.
Several back-end engines are implemeted in \golem:
\begin{itemize}
\item \texttt{lawi} is our re-implementation of the \textsc{Impact} algorithm~\cite{McMillan_2006}
\item \texttt{spacer} is our re-implementation of the \textsc{Spacer} algorithm~\cite{Komuravelli2016} and allows \golem{} to solve non-linear systems.
\item \texttt{tpa} is our new model-checking algorithm based on doubling abstractions using Craig interpolants~\cite{Blicha_2022, Blicha_2022b}.
\item \texttt{bmc} implements the standard algorithm of Bounded Model Checking~\cite{Biere_1999}
\item \texttt{kind} implements a basic variant of $k$-induction~\cite{Sheeran_2000}
\item \texttt{imc} is our implementation of McMillan's first interpolation-based model-checking algorithm~\cite{McMillan_2003}
\end{itemize}

\toolarchitecture

\golem{} is implemented in C++ and built on top of the interpolating SMT solver \textsc{OpenSMT}~\cite{OpenSMT2} which is used for both satisfiability solving and interpolation. The only dependencies are those inherited from  \textsc{OpenSMT}: Flex, Bison and GMP libraries.

\toolnew
Compared to the previous year, \golem{} has three new back-end engines: \texttt{bmc}, \texttt{kind} and \texttt{imc}.
However, these engines support only transition systems and did not participate in the competition for this reason.
Additionally, the preprocessing of the input system has improved significantly, without losing the ability to produce witnesses.

\toolconfiguration
For LIA-nonlin track we used only \texttt{spacer} engine; the other engines cannot handle nonlinear system yet.

\texttt{\$ golem --engine spacer}

For LIA-lin track, we used a trivial portfolio of \texttt{lawi}, \texttt{spacer} and \texttt{tpa} (in \texttt{split-tpa} mode) running independently.

\texttt{\$ golem --engine=spacer,lawi,split-tpa}

\toollink{https://github.com/usi-verification-and-security/golem}{MIT LICENSE}

\clearpage
%%%%% Edit below
\toolname{LoAT chc-comp-2023}

\def\CXX{{C\nolinebreak[4]\hspace{-.05em}\raisebox{.4ex}{\tiny\bf ++}}}

% People submitting the solver, and their affiliation.
% This does not necessarily include all tool authors.

\toolsubmitter{Florian Frohn}{LuFG Informatik 2, RWTH Aachen University, Germany}
\toolsubmitter{J\"urgen Giesl}{LuFG Informatik 2, RWTH Aachen University, Germany}

\toolalgorithm

The \emph{Loop Acceleration Tool} (LoAT) \cite{loat22} is based on
\emph{Acceleration Driven Clause Learning} (ADCL) \cite{adcl23}, a novel
calculus for analyzing satisfiability of CHCs. LoAT's implementation of ADCL is
based on a calculus for modular \emph{loop acceleration} \cite{accel20}. It can
analyze linear Horn clauses over integer arithmetic. While ADCL can also prove
satisfiability of CHCs, LoAT is currently restricted to proving
unsatisfiability. Besides unsatisfiability of CHCs, LoAT can also prove
non-termination and lower bounds on the worst-case runtime complexity of
transition systems.

\toolarchitecture

LoAT is implemented in \CXX. It uses the SMT solvers Z3~\cite{z308} and
Yices~\cite{yices14}, the recurrence solver PURRS~\cite{purrs}, and the automata
library libFAUDES~\cite{faudes}.

\toolnew

LoAT participates in the competition for the first time. Earlier version of LoAT
could not analyze CHCs, but only transition systems.

\toolconfiguration

At the competition, LoAT is run with the following arguments:
\begin{description}
\item[\protect{\tt --mode reachability}] for proving reachability for transition
systems or unsatisfiability of CHCs, respectively
\item[\protect{\tt --format horn}] for specifying that the input problem is
given in the SMT-LIB-format for Horn clauses
\end{description}

\toollink{https://loat-developers.github.io/LoAT/}{GPL licence}

\clearpage
%%%%% Edit below

\toolname{Theta v4.2.3}

% People submitting the solver, and their affiliation.
% This does not necessarily include all tool authors.

% I think this looks better than having the two-lined affiliation after each of our names, but feel free to change it if necessary.

% we use the orcidlink package to create clickable ORCID ID links, feel free to delete if unable to use. For arXiv, orcidlink has to be uploaded within the package, for example from here: https://raw.githubusercontent.com/duetosymmetry/orcidlink-LaTeX-command/master/orcidlink.sty

\toolsubmitter{M\'ark Somorjai\orcidlink{0000-0001-7537-0469}}{\vspace{-1\baselineskip}}
\toolsubmitter{Mih\'aly Dobos-Kov\'acs\orcidlink{0000-0002-0064-2965}}{\vspace{-1\baselineskip}}
\toolsubmitter{Levente Bajczi\orcidlink{0000-0002-6551-5860}}{\vspace{-1\baselineskip}}
\toolsubmitter{Andr\'as V\"or\"os\orcidlink{0000-0001-7617-3563}}{\\Department of Measurement and Information Systems\\Budapest University of Technology and Economics, Hungary}

\toolalgorithm
% 1--2 paragraphs describing the solver in general, including the
% algorithm/solving approach that is used. Include at most two key
% references to papers about the solver.

\textsc{Theta} decides the satisfiability of Constrained Horn Clauses by transforming it to a formal verification problem and employing an abstraction-based model checking technique. The input set of CHCs are transformed into a formal program representation named \emph{Control Flow Automata (CFA)} \cite{theta_cfa} in a way that the unsatisfiability of the CHC problem is equivalent to the reachability of erroneous locations in the CFA. A bottom-up transformation is used for linear CHCs while a top-down transformation is done to nonlinear CHCs \cite{theta_hcvs}. The erroneous state reachability of the created CFA is then checked using \emph{CounterExample-Guided Abstraction Refinement (CEGAR)} \cite{theta_cegar}, an iterative abstraction-based model checking algorithm.

\toolarchitecture
% 1--2 paragraphs describing the tool architecture and implementation,
% including important libraries and components that are used.

\textsc{Theta} is a highly configurable model checking framework implemented in Java \cite{theta_jar}. It supports various formalisms for the verification programs, engineering models and timed systems, among others. Verification is done by the main CEGAR engine, which utilizes SMT solvers through an SMTLIB interface to calculate interpolants and check the feasibility of paths. The CEGAR engine can be configured to use different abstraction domains and interpolation techniques. The framework offers a number of command line tools equipped with frontends that parse the input problem into a formalism. The bottom-up and top-down transformations from CHCs to CFA are implemented as a frontends for the \texttt{xcfa-cli} tool.

\toolconfiguration
% The configuration(s) of the solver used in the competition; if
% possible, specify commandline arguments to replicate the
% configuration.

\textsc{Theta} is run with a sequential portfolio of 3 configurations listed below, using explicit value tracking, split predicate or cartesian predicate abstraction. Interpolation was set to backwards binary interpolation or sequential interpolation, calculated by Z3 \footnote{\url{https://github.com/Z3Prover/z3}} as the underlying SMT solver.

\begin{enumerate}
    \item \texttt{--domain PRED\_SPLIT --refinement BW\_BIN\_ITP --predsplit WHOLE}
    \item \texttt{--domain PRED\_CART --refinement BW\_BIN\_ITP --predsplit WHOLE}
    \item \texttt{--domain EXPL --refinement SEQ\_ITP}
\end{enumerate}

\textsc{Theta} detects whether the input CHCs are linear or not and employs a bottom-up transformation for the former and a top-down transformation for the latter. The submitted Theta version and run scripts are available in the competition archive \cite{theta_archive}.

\toollink{https://github.com/ftsrg/theta}{Apache License 2.0}

\clearpage
% \newcommand{\toolname}[1]{\subsection{#1}}
% \newcommand{\toolsubmitter}[2]{\noindent #1\\#2\par\smallskip}
% \newcommand{\toolalgorithm}{\paragraph{Algorithm.}}
% \newcommand{\toolarchitecture}{\paragraph{Architecture and Implementation.}}
% \newcommand{\toolconfiguration}{\paragraph{Configuration in CHC-COMP-20.}}
% \newcommand{\toollink}[2]{\par\bigskip\noindent\url{#1}\\#2}

%%%%% Edit below

\toolname{Ultimate TreeAutomizer 0.2.3-dev-ac87e89}

% People submitting the solver, and their affiliation.
% This does not necessarily include all tool authors.

\toolsubmitter{Matthias Heizmann}{University of Freiburg, Germany}
\toolsubmitter{Daniel Dietsch}{University of Freiburg, Germany}
\toolsubmitter{Jochen Hoenicke}{University of Freiburg, Germany}
\toolsubmitter{Alexander Nutz}{University of Freiburg, Germany}
\toolsubmitter{Andreas Podelski}{University of Freiburg, Germany}
\toolsubmitter{Frank Schüssele}{University of Freiburg, Germany}

\toolalgorithm

The \textsc{Ultimate TreeAutomizer} solver implements an approach that is based on tree automata~\cite{TreeAutomizer}.
In this approach potential counterexamples to satisfiability are considered as a regular set of trees.
In an iterative \nobreak{CEGAR} loop we analyze potential counterexamples.
Real counterexamples lead to an \textit{unsat} result.
Spurious counterexamples are generalized to a regular set of spurious counterexamples
and subtracted from the set of potential counterexamples that have to be considered.
In case we detected that all potential counterexamples are spurious, the result is \textit{sat}.
The generalization above is based on tree interpolation and
regular sets of trees are represented as tree automata.

% 1--2 paragraphs describing the solver in general, including the
% algorithm/solving approach that is used. Include at most two key
% references to papers about the solver.

\toolarchitecture

% \textsc{Ultimate}
\textsc{TreeAutomizer} is a toolchain in the 
\textsc{Ultimate} framework.
This toolchain first parses the CHC input and then runs the \texttt{treeautomizer} plugin which
implements the above mentioned algorithm.
We obtain tree interpolants from the SMT solver SMTInterpol%
\footnote{\url{https://ultimate.informatik.uni-freiburg.de/smtinterpol/}}%
~\cite{cade/HoenickeS18}.
For checking satisfiability, we use the
and Z3 SMT solver%
\footnote{\url{https://github.com/Z3Prover/z3}}%
.
The tree automata are implemented in \textsc{Ultimate}'s automata library%
\footnote{\url{https://www.ultimate-pa.org/?ui=tool&tool=automata_library}}%
.
The \textsc{Ultimate} framework is written in Java and build upon the Eclipse Rich Client Platform (RCP). The source code is available at
GitHub\footnote{\url{https://github.com/ultimate-pa/}}.

% 1--2 paragraphs describing the tool architecture and implementation,
% including important libraries and components that are used.

\toolconfiguration

Our StarExec archive for the competition is shipped with the \texttt{bin/starexec\_run\_default}
shell script calls the \textsc{Ultimate} command line interface with the
\texttt{TreeAutomizer.xml} toolchain file and
the \texttt{TreeAutomizerHopcroftMinimization.epf} settings file.
Both files can be found in toolchain (resp. settings) folder of \textsc{Ultimate}'s repository.

% The configuration(s) of the solver used in the competition; if
% possible, specify commandline arguments to replicate the
% configuration.

\toollink{https://www.ultimate-pa.org/}{LGPLv3 with a linking exception for Eclipse RCP}

% Add bibtex entries for the references below

\clearpage
% \newcommand{\toolname}[1]{\subsection{#1}}
% \newcommand{\toolsubmitter}[2]{\noindent #1\\#2\par\smallskip}
% \newcommand{\toolalgorithm}{\paragraph{Algorithm.}}
% \newcommand{\toolarchitecture}{\paragraph{Architecture and Implementation.}}
% \newcommand{\toolconfiguration}{\paragraph{Configuration in CHC-COMP-20.}}
% \newcommand{\toollink}[2]{\par\bigskip\noindent\url{#1}\\#2}

%%%%% Edit below

\toolname{Ultimate Unihorn 0.2.3-dev-ac87e89}

% People submitting the solver, and their affiliation.
% This does not necessarily include all tool authors.

\toolsubmitter{Matthias Heizmann}{University of Freiburg, Germany}
\toolsubmitter{Daniel Dietsch}{University of Freiburg, Germany}
\toolsubmitter{Jochen Hoenicke}{University of Freiburg, Germany}
\toolsubmitter{Alexander Nutz}{University of Freiburg, Germany}
\toolsubmitter{Andreas Podelski}{University of Freiburg, Germany}
\toolsubmitter{Frank Schüssele}{University of Freiburg, Germany}

\toolalgorithm

\textsc{Ultimate Unihorn} reduces the satisfiability problem for a set of constraint Horn clauses
to a software verfication problem.
In a first step \textsc{Unihorn} applies a 
yet unpublished translation in which the constraint Horn clauses
are translated into a recursive program
that is nondeterministic and
whose correctness is specified by an assert statement
The program is correct (i.e., no execution violates the assert statement)
if and only if the set of CHCs is satisfiable.
For checking whether the recursive program satisfies its specification,
Unihorn uses \textsc{Ultimate Automizer}~\cite{tacas/HeizmannBDFHKNSSP23}
which implements an automata-based approach to software verification~\cite{cav/HeizmannHP13}.

% 1--2 paragraphs describing the solver in general, including the
% algorithm/solving approach that is used. Include at most two key
% references to papers about the solver.

\toolarchitecture

\textsc{Ultimate Unihorn} is a toolchain in the 
\textsc{Ultimate} framework.
This toolchain first parses the CHC input and then runs the \texttt{chctoboogie} plugin which
does the translation from CHCs into a recursive program.
We use the Boogie
% \footnote{\url{https://www.microsoft.com/en-us/research/project/boogie-an-intermediate-verification-language/}}
language to represent that program.
Afterwards the default toolchain for verifying a recursive Boogie programs by \textsc{Ultimate Automizer} is applied.
The \textsc{Ultimate} framework shares the libraries for handling SMT formulas with the SMTInterpol SMT solver.
While verifying a program, \textsc{Ultimate Automizer} needs SMT solvers
for checking satisfiability,
for computing Craig interpolants and
for computing unsatisfiable cores.
The version of \textsc{Unihorn} that participated in the competition
used the SMT solvers SMTInterpol%
\footnote{\url{https://ultimate.informatik.uni-freiburg.de/smtinterpol/}}%
and Z3%
\footnote{\url{https://github.com/Z3Prover/z3}}%
.
The \textsc{Ultimate} framework is written in Java and build upon the Eclipse Rich Client Platform (RCP). The source code is available at
GitHub\footnote{\url{https://github.com/ultimate-pa/}}.

% 1--2 paragraphs describing the tool architecture and implementation,
% including important libraries and components that are used.

\toolconfiguration

Our StarExec archive for the competition is shipped with the \texttt{bin/starexec\_run\_default}
shell script calls the \textsc{Ultimate} command line interface with the
\texttt{AutomizerCHC.xml} toolchain file and
the \texttt{chccomp-Unihorn\_Default.epf} settings file.
Both files can be found in toolchain (resp. settings) folder of \textsc{Ultimate}'s repository.

% The configuration(s) of the solver used in the competition; if
% possible, specify commandline arguments to replicate the
% configuration.

\toollink{https://www.ultimate-pa.org/}{LGPLv3 with a linking exception for Eclipse RCP}

% Add bibtex entries for the references below

%\input{solvers/spacer}

% ----------------------------------------------------------------
% \nocite{*}
\bibliographystyle{eptcs}
%\bibliography{biblio}

\clearpage
\appendix
\clearpage
\section{Detailed results}
\label{app:detres} 

\bigskip
% ---------------------------------------------------------------------------
\begin{table}[h]
\centering
\begin{tabular}{lrrrrrr}
\hline
\textbf{Solver}& \textbf{Score} & \textbf{\#sat} & \textbf{\#unsat} & 
\textbf{CPU time/s} & \textbf{Wall-clock/s} & \textbf{\#unique}\\\hline\hline
\rowcolor{lightgray} \spacer & 265 & 199 & 66 & 274397 & 138310 &  43\\\hline
\golem                       & 229 & 148 & 81 & 368980 & 129633 &   8\\\hline
\eldarica                    & 219 & 160 & 59 & 385851 & 112832 &  23\\\hline
\thetafix                    & 170 & 122 & 48 & 426006 & 370425 &   0\\\hline
\unihorn                     & 103 &  72 & 31 & 449683 & 384389 &   0\\\hline
\treeautomizer               & 81  &  50 & 31 & 537858 & 517349 &   0\\\hline
\loat                        & 50  &   0 & 50 & 287878 & 287841 &   4\\\hline
\end{tabular}
\caption{Solver performance on LIA-lin track}
\label{tab:res_lia_lin}
\end{table}

\bigskip
% ---------------------------------------------------------------------------
\begin{table}[h]
\centering
\begin{tabular}{lrrrrrr}
\hline
\textbf{Solver}& \textbf{Score} & \textbf{\#sat} & \textbf{\#unsat} & 
\textbf{CPU time/s} & \textbf{Wall-clock/s} & \textbf{\#unique}\\\hline\hline
\rowcolor{lightgray} \spacer & 384 & 235 & 149 &  90842 &  50781 & 38\\\hline
\eldarica                    & 330 & 185 & 145 & 218944 &  79522 &  9\\\hline
\golem                       & 310 & 178 & 132 & 248569 & 248578 &  3\\\hline
\unihorn                     & 121 &  72 &  49 & 470768 & 389915 &  0\\\hline
\thetafix                    &  38 &   8 &  30 & 687374 & 666145 &  0\\\hline
\treeautomizer               &  34 &   5 &  29 & 569895 & 531158 &  0\\\hline
\end{tabular}
\caption{Solver performance on LIA-nonlin track}
\label{tab:res_lia_nonlin}
\end{table}

\bigskip
% ---------------------------------------------------------------------------
\begin{table}[h]
\centering
\begin{tabular}{lrrrrrr}
\hline
\textbf{Solver}& \textbf{Score} & \textbf{\#sat} & \textbf{\#unsat} & 
\textbf{CPU time/s} & \textbf{Wall-clock/s} & \textbf{\#unique}\\\hline\hline
\rowcolor{lightgray} \spacer  & 281 & 212 & 69 & 359439 & 187454 & 81\\\hline
\eldarica                     & 220 & 150 & 70 & 478284 & 166185 & 15\\\hline
\thetafix                     & 184 & 134 & 50 & 285884 & 271624 &  0\\\hline
\unihorn                      & 164 & 122 & 42 & 242113 & 206799 &  1\\\hline
\treeautomizer                & 131 &  96 & 35 & 239591 & 229783 &  0\\\hline
\end{tabular}
\caption{Solver performance on LIA-lin-Arrays track}
\label{tab:res_lia_lin_arrays}
\end{table}

\clearpage
\bigskip
% ---------------------------------------------------------------------------
\begin{table}[h]
\centering
\begin{tabular}{lrrrrrr}
\hline
\textbf{Solver}& \textbf{Score} & \textbf{\#sat} & \textbf{\#unsat} & 
\textbf{CPU time/s} & \textbf{Wall-clock/s} & \textbf{\#unique}\\\hline\hline
\rowcolor{lightgray} \spacer & 258 & 148 & 110 & 290925 & 156914 & 75\\\hline
\eldarica                    & 206 & 122 &  84 & 454921 & 184851 & 26\\\hline
\unihorn                     &  96 &  37 &  59 & 234519 & 199416 &  0\\\hline
\thetafix                    &  85 &  45 &  40 & 588095 & 569760 &  4\\\hline
\treeautomizer               &  56 &   6 &  50 & 276025 & 250747 &  0\\\hline
\end{tabular}
\caption{Solver performance on LIA-nonlin-Arrays track}
\label{tab:res_lia_nonlin_arrays}
\end{table}

\bigskip
% ---------------------------------------------------------------------------
\begin{table}[h]
\centering
\begin{tabular}{lrrrrrr}
\hline
\textbf{Solver}& \textbf{Score} & \textbf{\#sat} & \textbf{\#unsat} & 
\textbf{CPU time/s} & \textbf{Wall-clock/s} & \textbf{\#unique}\\\hline\hline
\eldarica                   & 176 & 85 & 91 & 114521 &  42212 & 57\\\hline
\rowcolor{lightgray}\spacer & 120 & 59 & 61 & 195321 & 107046 &  1\\\hline
\end{tabular}
\caption{Solver performance on LIA-nonlin-Arrays-nonrecADT track}
\label{tab:res_lia_nonlin_Arrays_nonrecADT}
\end{table}

\bigskip
% ---------------------------------------------------------------------------
\begin{table}[h]
\centering
\begin{tabular}{lrrrrrr}
\hline
\textbf{Solver}& \textbf{Score} & \textbf{\#sat} & \textbf{\#unsat} & 
\textbf{CPU time/s} & \textbf{Wall-clock/s} & \textbf{\#unique}\\\hline\hline
\eldarica                   & 58 & 22 & 36 & 433561 & 150012 & 30\\\hline
\rowcolor{lightgray}\spacer & 30 &  3 & 27 & 440259 & 290358 &  2\\\hline
\end{tabular}
\label{tab:adt_nonlin}
\caption{Solvers performance on ADT-LIA-nonlin track}
\end{table}

\end{document}